\documentclass[12pt]{article}
\pdfoutput=1
\usepackage{xcolor}
\usepackage{braket}
\usepackage{bbm}
\usepackage[toc,page]{appendix}
\usepackage{mathtools}
\usepackage{hyperref}
\usepackage{textcomp}
\usepackage[DIV13]{typearea}
\usepackage{amsmath, amsthm, amssymb, mathtools,empheq,latexsym,dsfont}
\usepackage{bbm}
\usepackage{slashed, simplewick}
\usepackage[utf8]{inputenc}
\usepackage{graphicx,float,placeins}
\usepackage{makeidx}
\usepackage[font=small,labelfont=bf]{caption}
\usepackage{nicefrac}
\usepackage{subfigure}
\usepackage{array, bigdelim,multirow,multicol}
\usepackage{float}
\usepackage{rotating}
\usepackage{colortbl}
\usepackage{booktabs}
\usepackage[top=2cm,textwidth=16.6cm,textheight=22.75cm]{geometry}
\usepackage{doi}

\usepackage{gensymb}
\usepackage{makecell}
\usepackage{listings}
\usepackage{cancel}

\usepackage{cite}
\hypersetup{colorlinks=true,linkcolor=blue,citecolor=blue}

\graphicspath{{immagini/}}

\def\cW{{\cal W}} 
 
\def\cY{{\cal Y}}
\def\di{{\rm d}}
\def\1{\mathbf{1}}
\def\op{\mathbf{1'}}
\def\2{\mathbf{2}}
\def\3{\mathbf{3}}
\def\tp{\mathbf{3^\prime}}
\def\4{\mathbf{4}}
\def\5{\mathbf{5}}
\def\bc{\begin{center}}
\def\ec{\end{center}}
\def\dd{\displaystyle}

\newcommand{\be}{\begin{equation}}
\newcommand{\ee}{\end{equation}}
\newcommand{\bea}{\begin{eqnarray}}
\newcommand{\eea}{\end{eqnarray}}

\begin{document}
 \unitlength = 1mm

\setlength{\extrarowheight}{0.2 cm}

\thispagestyle{empty}

\bigskip

\vskip 1cm

\begin{center}
\vspace{1.5cm}
  {\Large\bf Modular Invariant Models of Lepton Masses \\[.3cm]
  at levels 4 and 5}\\[1.3cm]
    {\bf Juan Carlos Criado$^{a}$, Ferruccio Feruglio$^{b}$, Simon J.D. King $^{c,b}$}     \\[0.5cm]
    {\em $(a)$  CAFPE and Departamento de F\'isica Te\'orica y del Cosmos, Universidad de Granada, Campus de
Fuentenueva, E-18071, Granada, Spain}\\   
     {\em $(b)$  Dipartimento di Fisica e Astronomia `G.~Galilei', Universit\`a di Padova\\
INFN, Sezione di Padova, Via Marzolo~8, I-35131 Padua, Italy} \\
{\em $(c)$ School of Physics and Astronomy, University of Southampton\\ Southampton, SO17 1BJ, United Kingdom}
 \\[1.0cm]
\end{center}

\centerline{\large\bf Abstract}
\begin{quote}
\indent
We explore alternative descriptions of the charged lepton sector in modular invariant models of lepton masses and mixing angles.
In addition to the modulus, the symmetry breaking sector of our models includes ordinary flavons. Neutrino mass
terms depend only on the modulus and are tailored to minimize the number of free parameters. The charged lepton Yukawa couplings
rely upon the flavons alone. We build modular invariant models at levels 4 and 5, where neutrino masses are described both
in terms of the Weinberg operator or through a type I seesaw mechanism. At level 4, our models reproduce the hierarchy among electron, muon and tau masses by letting the weights play the role of Froggatt-Nielsen charges. At level 5, our setup allows the treatment of left and right handed charged leptons
on the same footing. We have optimized the free parameters of our models in order to match the
experimental data, obtaining a good degree of compatibility and predictions for the absolute neutrino masses and the $CP$ violating phases.
At a more fundamental level, the whole lepton sector could be
correctly described by the simultaneous presence of several moduli. Our examples are meant to make a first step in this direction.
\end{quote}

\newpage 

\section{Introduction}
Masses and mixing angles of elementary fermions are known with good precision and in the last few years the progress in the lepton sector
has been particularly impressive, with neutrino squared mass differences and mixing angles that are attaining or approaching  percent-level precision. 
Despite such an advance on the experimental side,
the fundamental principle, if any, ruling this important aspect of fundamental interactions has remained elusive. In recent times a lot of attention has been focused
on neutrinos, since the relatively mild mass hierarchy and the large mixing angles discovered through neutrino oscillations have not matched the expectations
based on the knowledge of the quark sector. Neutrinos led to a change of perspective, particularly relevant when we look at the  flavour puzzle in the light of a unified theory, where leptons and quarks loose their individuality. 

One of the few tools we have to address the flavour puzzle is the one 
based on flavour symmetries, which, however, comes with its own drawbacks. Flavour symmetries cannot be exact symmetries \cite{Reyimuaji:2018xvs} and Yukawa couplings 
are usually expressed as a power series in the symmetry breaking terms, with many independent free variables, to the detriment of predictability.
In addition, such an approach typically makes use of several symmetry breaking parameters, with specific orientation in  flavour space, considerably
complicating the construction. Finally, the most popular flavour symmetries of the lepton sector constrain only mixing angles and phases, leaving fermion
masses essentially undetermined \cite{Altarelli:2010gt,Ishimori:2010au,Hernandez:2012ra,King:2013eh,King:2014nza,Feruglio:2015jfa,King:2017guk,Hagedorn:2017zks}.

           Recently, modular invariance has been invoked as candidate flavour symmetry \cite{Feruglio:2017spp}. In its simplest implementation a unique complex field, the modulus,
acts as symmetry breaking parameter, thus simplifying the vacuum alignment problem. Modular invariance, in the limit of exact supersymmetry, completely
determines the Yukawa couplings, to any order of the expansion in powers of the modulus. Moreover, neutrino masses, mixing angles and phases are all related to each other
and, in minimal models, depend only on a few parameters. The formalism has been extended to consistently include $CP$ transformations \cite{Novichkov:2019sqv} \footnote{The interplay between $CP$ and modular invariance in string theory have been discussed in Ref. \cite{Dent:2001cc,Dent:2001mn} and especially in Ref. \cite{Baur:2019kwi,Baur:2019iai} where a unified picture of flavour, $CP$ and modular invariance has been analyzed from a string theory perspective.} and it can involve several moduli   \cite{Ferrara:1989qb,deMedeirosVarzielas:2019cyj}. 
The idea that Yukawa couplings are determined by a set of moduli is clearly not new, and has been naturally realized in the context of string theory
\cite{Hamidi:1986vh,Dixon:1986qv,Lauer:1989ax,Lauer:1990tm,Erler:1991nr}, in D-brane compactification \cite{Cremades:2003qj,Blumenhagen:2005mu,Abel:2006yk,Blumenhagen:2006ci,Marchesano:2007de,Antoniadis:2009bg,Kobayashi:2016ovu}, in magnetized extra dimensions \cite{Cremades:2004wa,Abe:2009vi,Kobayashi:2018rad}, and in orbifold compactification \cite{Ibanez:1986ka,Casas:1991ac,Lebedev:2001qg,Kobayashi:2003vi}. Modular invariance has also been incorporated in early
flavour models \cite{Brax:1994kv,Binetruy:1995nt,Dudas:1995eq,Dudas:1996aa,Leontaris:1997vw}.
However, the 
main advantage of the recent approach is that it can be implemented in a bottom-up perspective, relying on the group transformation properties of
modular forms of given weight and level. 

          Several models of lepton masses and mixing angles have been built at level 2 \cite{Kobayashi:2018vbk,Kobayashi:2018wkl}, 3 \cite{Feruglio:2017spp,Criado:2018thu,Kobayashi:2018scp,Novichkov:2018yse,Ding:2019zxk}, 4 \cite{Penedo:2018nmg,Novichkov:2018ovf,King:2019vhv} and 5 \cite{Novichkov:2018nkm,Ding:2019xna}. Extensions to quarks \cite{Okada:2018yrn,Okada:2019uoy} and to grand unified theories  \cite{deAnda:2018ecu,Kobayashi:2019rzp} have also been proposed. In most of the existing constructions, there is a unique symmetry breaking parameter: the 
modulus itself. While this scenario is certainly appealing since it minimizes the symmetry breaking sector, it does not yet provide a convincing explanation of the 
charged lepton masses. The mass hierarchy is achieved by hand by introducing one parameter for each charged lepton species.
This can be intuitively understood by recognizing that the dependence of modular forms on the modulus is nearly exponential and small neutrino mass hierarchies and large mixing angles 
require a modulus with small imaginary part, which is inadequate to generate the large hierarchies observed among electron, muon and tau masses.
This may indicate that the charged lepton sector requires a different description, perhaps in terms of more moduli, a natural possibility in string theory.

In the present work we explore alternative descriptions of the charged lepton sector in a modular invariant framework. We test the dependence of charged lepton masses
on an additional set of fields by including in the symmetry breaking sector both the modulus and ordinary flavons, chiral multiplets
invariant under gauge transformations carrying non-trivial representations of the finite modular groups and non-trivial weights, to guarantee consistence with invariance
under the full modular group. This has been done at level 3 in Ref. \cite{Criado:2018thu} and at level 5 in Ref. \cite{Ding:2019xna}. We will extend the investigation to level 4 and extend  the possibilities studied so far al level 5. At level 4 the charged lepton Yukawa couplings
are tailored to depend only on the flavons, with the hope of reproducing charged lepton masses with parameters similar in size, at least at the level of order of magnitudes.
We will let right-handed charged leptons be responsible for the observed mass hierarchy, by assigning them different modular weights
compensated by growing powers of the flavons, much as in Ref. \cite{Dudas:1995eq,Dudas:1996aa,Leontaris:1997vw}. At level 5 we will take a more radical departure
from the existing constructions and we will assign the right-handed charged leptons to an irreducible triplet of $\Gamma_5$, to treat them more closely to their left-handed partners.
In our models only the neutrino sector depends non-trivially on the modulus. As done in Ref. \cite{Criado:2018thu}, we will not attempt to dynamically select the vacuum
configurations in the symmetry breaking sector. We have no compelling indications so far that Nature follows a dynamical principle to set the cosmological constant or the
electroweak scale. We thus treat the vacuum expectation values (VEVs) as free parameters, to be varied to match the experimental data. 

The models are built aiming at minimizing the number of free parameters. So far few predictive models use four independent parameters to describe neutrino masses, mixing angles and phases and a variety
of models achieve that with five free parameters, including real and imaginary part of the modulus. As we will see the models we have been able to construct
make use of at least five parameters and can be considered next-to-minimal. In our attempts we have also incorporated $CP$ invariance, to be spontaneously
broken by the modulus and by the flavons. We present realistic examples where neutrino masse are described both in terms of the Weinberg operator and via the type I seesaw mechanism.

        Our paper is organized as follows. In section 2 we briefly review the formalism of modular invariant supersymmetric theories applied to the lepton sector and we will 
describe our models. In section 3 we present the data, describe our fit and we show the results of the fit and the predictions of the models. Finally
in   4 we draw our conclusion. 
\section{The Models}
We brefly review the formalism of modular invariant supersymmetric theories \cite{Ferrara:1989bc,Ferrara:1989qb}.
The models analyzed here are supersymmetric and gauge invariant under SU(3)$\times$SU(2)$\times$U(1).
We are mainly interested to the Yukawa interactions, described by the action:
\be
{\cal S}=\int d^4 x d^2\theta d^2\bar \theta~ K(\Phi,\bar \Phi)+\int d^4 x d^2\theta~ w(\Phi)+h.c.~~~,
\label{action}
\ee
where $K(\Phi,\bar\Phi)$, the K\"ahler potential, is a real gauge-invariant function of the chiral superfields $\Phi$ and their conjugates
and $w(\Phi)$, the superpotential, is a holomorphic gauge-invariant function of the chiral superfields $\Phi$.
The chiral superfields $\Phi=(\tau,\varphi^{(I)})$ include the modulus $\tau$, a dimensionless chiral supermultiplet, and the remaining chiral supermultiplets, $\varphi^{(I)}$. Under the modular group $\Gamma$ the modulus transforms as 
\be
\tau\to \gamma\tau\equiv\dd\frac{a \tau+b}{c \tau+d}~~~,
\label{ttau}
\ee
with $a$, $b$, $c$ and $d$ integers satisfying $ad-bc=1$. The modular group $\Gamma$ is an infinite discrete group, generated by the elements $S$ and $T$ satisfying
$S^2=(ST)^3=1$. They act as
\be
\tau\to -\frac{1}{\tau}~~~~~~~~~~~~~(S)~~~~~~~~~~~~~~~~~~~~\tau\to\tau+1~~~~~~~~~~~~~(T)~~~.
\ee
The transformation properties of $\varphi^{(I)}$ are fully specified by the data $(k_I,N,\rho^{(I)})$, where
$k_I$ (the weight) is a real number, $N$ (the level) is an integer and $\rho^{(I)}$ is a unitary representation of the quotient group $\Gamma_N=\Gamma/\Gamma(N)$. $\Gamma(N)$ is a principal congruence subgroup of $\Gamma$ and the level $N$ can be kept fixed in the construction. The multiplets $\varphi^{(I)}$ transform as
\be
\varphi^{(I)}\to (c\tau+d)^{k_I} \rho^{(I)}(\gamma) \varphi^{(I)}~~~.
\label{tphi}
\ee
We choose a minimal form of the Kahler potential, invariant under (\ref{ttau},\ref{tphi}) up to Kahler transformations:
\be
K(\Phi,\bar \Phi)=-h \log(-i\tau+i\bar\tau)+ \sum_I (-i\tau+i\bar\tau)^{-k_I} |\varphi^{(I)}|^2~~~,
\label{kalex}
\ee
where $h$ is a positive constant.
Concerning the superpotential $w(\Phi)$, its expansion in power series of the supermultiplets $\varphi^{(I)}$ reads:
\be
w(\Phi)=\sum_n Y_{I_1...I_n}(\tau)~ \varphi^{(I_1)}... \varphi^{(I_n)}~~~.
\label{psex}
\ee
For the $n$-th order term to be modular invariant the functions $Y_{I_1...I_n}(\tau)$ should be modular forms of weight $k_Y(n)$ and level $N$, transforming in the representation $\rho$ of $\Gamma_N$:
\be
Y_{I_1...I_n}(\gamma\tau)=(c\tau+d)^{k_Y(n)} \rho(\gamma)~Y_{I_1...I_n}(\tau)~~~,
\ee
satisfying the conditions:
\begin{enumerate}
\item[1.]
The weight $k_Y(n)$ should compensate the overall weight of the product $\varphi^{(I_1)}... \varphi^{(I_n)}$:
\be
k_Y(n)+k_{I_1}+....+k_{I_n}=0~~~.
\label{compensate}
\ee
\item[2.]
The product $\rho\times \rho^{{I_1}}\times ... \times \rho^{{I_n}}$ contains an invariant singlet.
\end{enumerate}
The above requirement is very restrictive.
Indeed, for each level $N$ and for each even non-negative weight $k$, there is only a finite number of linearly independent modular forms
\footnote{Recently modular forms of general integer weights and their transformation properties under the double covering of finite modular groups have been analyzed in Ref. \cite{Liu:2019khw}.}. They span the linear space ${\cal M}_k(\Gamma(N))$. Forms with vanishing weight are constant, that is independent from $\tau$. We will analyze models
with $N=4$ and 5. The dimension of ${\cal M}_k(\Gamma(4))$ is $2k+1$, while 
${\cal M}_k(\Gamma(5))$ has dimension $5k+1$. Modular forms of weight 2 generate the whole ring of modular forms. The five independent
modular forms of level 4 and weight 2 have been constructed in Ref. \cite{Penedo:2018nmg}. They decompose as $\2+\tp$ under the finite group $\Gamma_4\equiv S_4$. The eleven independent
modular forms of level 5 and weight 2 have been constructed in Ref. \cite{Novichkov:2018nkm} and \cite{Ding:2019xna}. They decompose as $\3+\tp+\5$ under $\Gamma_5\equiv A_5$. In Appendix \ref{appendix:level_4} and \ref{app5} we list them. 

The chiral multiplets
$\varphi^{(I)}$ comprise three generations of lepton singlets $E^c$ and doublets $L$,
the Higgses $H_{u,d}$, and gauge invariant flavons $\varphi$. We will consider both the case where neutrino masses arise
through the Weinberg operator and the case where neutrinos get their masses through the seesaw
mechanism. In the latter framework also three generations of gauge singlets $N^c$ are included. In our conventions both the modulus $\tau$ and the flavon $\varphi$ are dimensionless fields. The correct dimensions can be recovered by an appropriate rescaling. 

Invariance under $CP$ can be incorporated in a consistent way \cite{Novichkov:2019sqv} by requiring:
\be
\tau \xrightarrow{\text{$CP$}} -\tau^*~~~,
\label{tauCP}
\ee
up to a modular transformation. On the chiral multiplets $\varphi^{(I)}$ a $CP$ transformation acts as
\be
\varphi^{(I)} \xrightarrow{\text{$CP$}} X_{(I)} [\varphi^{(I)}]^*~~~,
\ee
where $X_{(I)}$ is a matrix satisfying the consistency conditions:
\be
X_{(I)} [\rho^{(I)}(\gamma)]^* X_{(I)}^{-1}= \rho^{(I)}(\gamma')~~~,~~~~~~~~~~~(\gamma,\gamma')\in \Gamma~~~.
\ee
In a basis where all the matrices $\rho^{(I)}(\gamma)$ are symmetric, these conditions are always solved by $X_{(I)}=\mathbbm{1}$. This is the case of our choice of basis at level 5.
At level 4 our basis does not enjoy this property and a non-canonical solution for $X_{(I)}$ is listed in Appendix A.
\subsection{Level 4 models}
The group $\Gamma_4$ has order 24 and is isomorphic to $S_4$. Its irreducible representations are
$\1$, $\op$, $\2$, $\3$ and $\tp$. It is generated by two elements $S$ and $T$
satisfying the relations $S^2 \,=\, (ST)^3 \,=\, T^4 \,=\, \mathds{1}$.
In Appendix \ref{appendix:level_4} {we detail the explicit form of the generators for the irreducible representations and the relevant Clebsch-Gordan coefficients used in this paper.}
The particle content, weights and representations {of our models} are shown in Tab. \ref{tabmod3w}.

\begin{table}[h!] 
\centering
\begin{tabular}{|c|c|c|c|c|c|c||c|c|}
\hline
&$E_1^c$&$E_2^c$&$E_3^c$&$N^c$& $L$& $H_{u,d}$&$\varphi$&$\varphi'$\rule[-2ex]{0pt}{5ex}\\
\hline
$\Gamma_4\equiv S_4$&$\1$&$\1$&$\1$& $\3$& $\3$& $\1$& $\3$& $\op$ \rule[-2ex]{0pt}{5ex}\\
\hline
$k_I$ ({\tt Seesaw})&$k-3k_\varphi$&$k-2k_\varphi$&$k-k_\varphi$& $k$& $-k$& $0$& $k_\varphi$ & $k_{\varphi'}$ \rule[-2ex]{0pt}{5ex}\\
\hline
$k_I$ ({\tt Weinberg})&$-k-3k_\varphi$&$-k-2k_\varphi$&$-k-k_\varphi$& $-$& $k$& $0$& $k_\varphi$& $k_{\varphi'}$ \rule[-2ex]{0pt}{5ex}\\
\hline
\end{tabular}
\caption{{Chiral supermultiplets, transformation properties and weights. Weights for $E_i^c$ and $L$ depend
on whether neutrinos get their masses from the seesaw mechanism or from the Weinberg operator. A possible choice leading to the superpotential given in the text is $k=-5/3$, $k_{\varphi'}=+4/3$ and $k_{\varphi}=+3/2$. As a consequence,
the neutrino sector depends only on $\varphi'$ and the charged lepton sector depends only on $\varphi$.}}
\label{tabmod3w}
\end{table}

\noindent
With the above assignment the superpotential reads
\be
w=w_h+w_e+w_\nu~~~,
\label{sp1}
\ee
where $w_h$, $w_e$, $w_\nu$ describe the Higgs sector, the charged lepton sector and the neutrino sector, respectively.
Since the Higgs sector plays no role in our discussion, we neglect $w_h$. We set $H_u=H_d=1$ {in the superpotential,} but we keep track of the correct dimension of the operators. 

In the neutrino sector $w_\nu$ depends on the mass generation mechanism.
When neutrino masses originate from the Weinberg operator we have:
\be
w_\nu=-\dd\frac{1}{\Lambda}\left[(\varphi' L L~ Y_{\bf 2})_1+ \xi (\varphi'L L~  Y_{\bf 3'})_1\right]~~~,
\label{weinb1}
\ee
where $\Lambda$ stands for the scale associated to lepton number violation, $(...)_r$ denotes the $r$ representation of $\Gamma_4$ and $\xi$ is a {free parameter.} When light neutrinos get their masses from the seesaw mechanism, the terms of $w_\nu$ bilinear in the matter multiplets $L$ and $N^c$ read
\be
w_\nu=-y_0(N^c L)_1+ \Lambda \left[(\varphi' N^cN^c~ Y_{\bf 2})_1+ \xi (\varphi'N^cN^c~ Y_{\bf 3'})_1\right]+... 
\ee
Dots denote terms containing three or more powers of the matter fields, having no impact on our analysis. 
{A truly minimal model would involve a single invariant in the neutrino sector. For instance, a suitable assignement of weights can allow the unique term $w_\nu=-(L L~ Y_{\bf 2})_1/\Lambda$ (Weinberg) or $w_\nu=-(N^c N^c~ Y_{\bf 2})_1 \Lambda$ (seesaw). We have studied these possibilities, but we found no viable choice of parameters which may reproduce data.}

At energies below the mass scale $\Lambda$ for both models we have, in a matrix notation:
\be
w_\nu= -\dd\frac{1}{\Lambda}L^T {\cal W} L+...~~~,
\ee
where ${\cal W}$ denotes a matrix in generation space depending on the 5 independent level 4 and weight +2 modular forms $Y_i(\tau)$ $(i=1,...,5)$.  We list these results in table \ref{tab:S4-W-matrix}, where the VEV of $\varphi'$ has been absorbed in $\Lambda$, $Y_{i}$ stands for $Y_{i}(\tau)$, and the indices $W,S$ distinguish neutrino masses originating from the Weinberg operator or from the seesaw mechanism.

\begin{table}[h!]
  \centering
  \begin{tabular}{cc}
    \toprule
    \texttt{Weinberg}, &
    $ {\cal W}_W=
    \left(
 \begin{array}{ccc}
0&Y_1&- Y_2\\
Y_1&- Y_2&0\\
- Y_2&0&Y_1
 \end{array}
 \right)
 +\xi
 \left(
 \begin{array}{ccc}
2 Y_3&-Y_5&-Y_4\\
-Y_5&2Y_4&-Y_3\\
-Y_4&-Y_3&2Y_5
 \end{array}
 \right)$
    \\
    \midrule
    \texttt{Seesaw},  &
${\cal W}_S=
\dd\frac{y_0^2}{2}    
\left(
 \begin{array}{ccc}
1&0&0\\
0&0&1\\
0&1&0
 \end{array}
 \right) {\cal W}_W^{-1}
 \left(
 \begin{array}{ccc}
1&0&0\\
0&0&1\\
0&1&0
 \end{array}
 \right)$
    \\
    \bottomrule
  \end{tabular}
  \caption{{Relevant matrices in the neutrino sector of the superpotential in $\Gamma_4$
    models.}}
  \label{tab:S4-W-matrix}
\end{table}

\noindent
The light neutrino mass matrix $m_\nu$ is
\be
m_\nu={\cal W}\frac{v^2}{\Lambda}\sin^2\hat\beta~~~,
\label{eq:mnu}
\ee
\vskip 0.2 cm
\noindent
{where $\tan\hat\beta$ is the ratio of VEVs, $\braket{H_u}/\braket{H_d}$}.
So far, the results in the neutrino sector would not vary had we instead defined $N^{c}$ and $L$ to transform as a $\textbf{3'}$, rather than a $\textbf{3}$ under $\Gamma_4$. However, the following discussion in the charged lepton sector requires the properties as defined in Tab. \ref{tabmod3w}. The superpotential $w_e$ for the charged lepton sector reads:
\be
\label{we}
w_e=-a E_1^c (L~\varphi^3)_1-a' E_1^c (L~\varphi^3)'_1-b E_2^c (L~\varphi^2)_1-cE_3^c (L~\varphi)_{1}\equiv-{E^c}^T {\cal Y}_e L~~~.
\ee
In the last equality we use a vector notation and
\be
{\cal Y}_e=
\left(
 \begin{array}{ccc}
a(\varphi_2^3-2\varphi_1^3+\varphi_3^3)&3a (\varphi_1\varphi_2^2-\varphi_2\varphi_3^2)&-3a(\varphi_2^2\varphi_3-\varphi_1\varphi_3^2)\\
+a'(\varphi_1^3+2\varphi_1\varphi_2\varphi_3)&+a'(\varphi_1^2\varphi_3+2\varphi_2\varphi_3^2)&+a'(\varphi_1^2\varphi_2+2\varphi_2^2\varphi_3)\rule[-2ex]{0pt}{5ex}\\
b(\varphi_1^2-\varphi_2\varphi_3)&b(\varphi_2^2-\varphi_1\varphi_3)&b(-\varphi_1\varphi_2+\varphi_3^2)\rule[-2ex]{0pt}{5ex}\\
c\varphi_1&c\varphi_3&c\varphi_2
 \end{array}
 \right)~~~.
 \label{clept}
\ee
There are two independent $\Gamma_4$ invariants that can be built out of $L$ and $\varphi^3$, hence the two independent parameters
$a$ and $a'$.
The dependence on the flavon supermultiplet $\varphi$ is fixed by the weight assignment.
There is no dependence on the modulus $\tau$, since the bilinears $(E_1^c L,E_2^c L,E_3^c L)$ have weight $(-3k_\varphi,-2k_\varphi,-k_\varphi)$. Taking, for instance, $k_\varphi=+3/2$, these weights cannot be matched by modular forms.
The charged lepton mass matrix $m_e$ reads
\be
m_e= {\cal Y}_e\frac{v}{\sqrt{2}} \cos\hat\beta~~~.
\label{me}
\ee
{Notice that if the flavon $\varphi$ is aligned along the $(0,\varphi_2,0)$ direction, ${\cal Y}_e$ is diagonal and the charged lepton masses are given by:
\be
m_e= \frac{a}{\sqrt{2}}v \varphi_2^3 \cos\hat\beta~~~,~~~~~~~
m_\mu= \frac{b}{\sqrt{2}}v \varphi_2^2 \cos\hat\beta~~~,~~~~~~~
m_\tau= \frac{c}{\sqrt{2}}v \varphi_2 \cos\hat\beta~~~.
\ee
Hence, a mass hierarchy can be generated by $|\varphi_2|<1$, even with $a$, $b$ and $c$ of the same order.

In our numerical analysis we will treat the modulus $\tau$ and the VEV of $\varphi$ as free parameters. Beyond that, the parameters controlling lepton masses and mixing angles are the overall scale $\Lambda$ and the five dimensionless constants $\xi$, $a$, $a'$, $b$ and $c$. Without loss of generality, we can require
$a$, $a'$, $b$ and $c$ to be real, since their phases are always unphysical. On the contrary, the phase of $\xi$
cannot be removed by a field redefinition. We will consider two options, either requiring the theory to be invariant under $CP$ at the Lagrangian level, or not. In the former case, using the $CP$ transformation given in Appendix A, we find that $\xi$ should be real and $CP$ can be spontaneously broken by the VEVs of $\tau$ and/or $\varphi$. In the latter case, we will treat $\xi$ as a complex free parameter.
The dependence on $\tan\hat\beta$ can be absorbed into the above parameters and will not be explicitly shown when reporting numerical values.}

\subsection{Level 5 models}

The irreducible representations of the group $\Gamma_5 \equiv A_5$ are $\mathbf 1$, $\mathbf 3$,
$\mathbf 3'$, $\mathbf 4$ and $\mathbf 5$. Its generators are $S$ and $T$, satisfying
$S^2 = {(ST)}^3 = T^5 = \mathds{1}$. In appendix~\ref{app5} we
{specify} the explicit form of the generators for each representation, together
with the relevant Clebsh-Gordan coefficients.
Here, we construct modular-invariant models in which all leptons are collected
into $\3$ or $\tp$ multiplets of $A_5$, containing the three generations of each
type of field. We take the neutrino sector to be minimal, it should
only depend on the modulus $\tau$ and an overall scale. Modular forms will not
appear in the charged-lepton sector, which instead will contain two extra
flavons. In table \ref{tab:A5-models}, we show the assignments of representations
and weights that we consider.

\begin{table}[h] 
  \centering
  \begin{tabular}{|c|c|c|c|c||c|c|}
    \hline
    & $E^c$ & $N^c$ & $L$ & $H_{u,d}$ & $\varphi$ & $\chi$ \\
    \hline
    {$\Gamma_5\equiv A_5$} & $\rho_L$ & $\rho_N$ & $\rho_L$ & 1 & $\rho_L$ & 1 \\
    \hline
    $k_I$ & $-3-k_L$ & $k_N$ & $k_L$ & 0 & $3/2$ & $3/2$ \\
    \hline
  \end{tabular}
  ~
  \begin{tabular}{|c|cc|cc|}
    \hline
    & $\rho_N$ & $\rho_L$ & $k_N$ & $k_L$ \\
    \hline
    \multirow{2}{*}{\texttt{Weinberg}} & -- & $\3$ & -- & -1 \\
    & -- & $\tp$ & -- & -1 \\
    \hline
    \multirow{4}{*}{\texttt{Seesaw}} & $\3$ & $\3$ & -1 & 1 \\
    & $\tp$ & $\tp$ & -1 & 1 \\
    & $\3$ & $\tp$ & 0 & -2 \\
    & $\tp$ & $\3$ & 0 & -2 \\
    \hline
  \end{tabular}
  \caption{Chiral supermultiplets, transformation properties and weights for the
    level-5 models.}
  \label{tab:A5-models}
\end{table}

Setting $H_u = H_d = 1$, the neutrino sector $w_\nu$ of the superpotential is,
depending on the choice Weinberg vs. Seesaw and $(\rho_L =
\rho_N)$ vs. $(\rho_L \neq \rho_N)$:
\begin{equation}
  w_\nu =
  \left\{
    \begin{array}{ll}
      -\frac{1}{\Lambda} {\left(LL \, Y_5\right)}_1 &
      \mathtt{Weinberg} \\
      -y_0 (N^c L)_1 + \Lambda {\left(N^c N^c \, Y_\5\right)}_1 &
      \mathtt{Seesaw}, \; \rho_L = \rho_N \\
      -y_0 (N^c L \, Y_\5)_1 + \Lambda {\left(N^c N^c\right)}_1 &
      \mathtt{Seesaw}, \; \rho_L \neq \rho_N
    \end{array}
  \right.
\end{equation}
The case of $\rho_L \sim \rho_N \sim 3$ has been studied in detail in Ref. \cite{Ding:2011cm} and so not discussed here.

Below the energy scale $\Lambda$, $w_\nu$ can always be written as
\begin{equation}
  w_\nu = -\frac{1}{\Lambda} L^T \cW L,
\end{equation}
with $\cW$ a $3 \times 3$ matrix, whose explicit form for each case can be read
from table~\ref{tab:A5-W-matrix}, using the equation
$\cW = \frac{y_0^2}{2} \, \cY_\nu^T \cW_W^{-1} \cY_\nu$ for the seesaw case. The
light neutrino mass matrix $m_\nu$ can be obtained from $\cW$ as in
Eq.~\ref{eq:mnu}.
\begin{table}[h]
  \centering
  \begin{tabular}{cc}
    \toprule
    $\begin{array}{c}
    \texttt{Weinberg}, ~\rho_L = \3 \\
    k_L=-1
    \end{array}$
    &
    $ \cW = \left(\begin{array}{ccc}
        2 Y_1 & -\sqrt{3} Y_5 & -\sqrt{3} Y_2 \\
        -\sqrt{3} Y_5 & \sqrt{6} Y_4 & -Y_1 \\
        -\sqrt{3} Y_2 & -Y_1 & \sqrt{6} Y_3
      \end{array}\right) $
    \\
    \midrule
    $\begin{array}{c}
    \texttt{Weinberg}, ~\rho_L = \tp \\
    k_L = -1
    \end{array}$
    &
    $\cW = \left(\begin{array}{ccc}
        2 Y_1 & -\sqrt{3} Y_4 & -\sqrt{3} Y_3 \\
        -\sqrt{3} Y_4 & \sqrt{6} Y_2 & -Y_1 \\
        -\sqrt{3} Y_3 & -Y_1 & \sqrt{6} Y_5
      \end{array}\right)$
    \\
    \midrule
    $\begin{array}{c}
    \texttt{Seesaw}, ~\rho_L = \rho_N = \3\\
    k_L=1 ,~ k_N=-1
    \end{array} $ 
    &
    $\cW_W = \left(\begin{array}{ccc}
       2 Y_1 & -\sqrt{3} Y_5 & -\sqrt{3} Y_2 \\
        -\sqrt{3} Y_5 & \sqrt{6} Y_4 & -Y_1 \\
        -\sqrt{3} Y_2 & -Y_1 & \sqrt{6} Y_3
      \end{array}\right)$,
    $\cY_\nu = \left(\begin{array}{ccc}
        1 & 0 & 0 \\ 0 & 0 & 1 \\ 0 & 1 & 0
      \end{array}\right)$
    \\
    \midrule
    $\begin{array}{c}
    \texttt{Seesaw}, ~\rho_L = \rho_N = \tp \\
    k_L=1 ,~ k_N=-1
    \end{array} $ 
     &
    $\cW_W = \left(\begin{array}{ccc}
        2 Y_1 & -\sqrt{3} Y_4 & -\sqrt{3} Y_3 \\
        -\sqrt{3} Y_4 & \sqrt{6} Y_2 & -Y_1 \\
        -\sqrt{3} Y_3 & -Y_1 & \sqrt{6} Y_5
      \end{array}\right)$,
    $\cY_\nu = \left(\begin{array}{ccc}
        1 & 0 & 0 \\ 0 & 0 & 1 \\ 0 & 1 & 0
      \end{array}\right)$
    \\
        \midrule
        $\begin{array}{c}
 \texttt{Seesaw}, \rho_L = 3, \rho_N = \tp \\
    k_L=-2 ,~ k_N=0
    \end{array} $ 
    &
    $\cW_W = \left(\begin{array}{ccc}
        1 & 0 & 0 \\ 0 & 0 & 1 \\ 0 & 1 & 0
      \end{array}\right)$,
    $\cY_\nu = \left(\begin{array}{ccc}
        \sqrt{3} Y_1 & Y_4 & Y_3 \\
        Y_5 & -\sqrt{2} Y_3 & -\sqrt{2} Y_2 \\
        Y_2 & -\sqrt{2} Y_5 & -\sqrt{2} Y_4
      \end{array}\right)$
    \\
    \midrule
        $\begin{array}{c}
 \texttt{Seesaw}, \rho_L = 3', \rho_N = \3 \\
    k_L=-2 ,~ k_N=0
    \end{array} $  &
    $\cW_W = \left(\begin{array}{ccc}
        1 & 0 & 0 \\ 0 & 0 & 1 \\ 0 & 1 & 0
      \end{array}\right)$,
    $\cY_\nu = \left(\begin{array}{ccc}
        \sqrt{3} Y_1 & Y_5 & Y_2 \\
        Y_4 & -\sqrt{2} Y_3 & -\sqrt{2} Y_5 \\
        Y_3 & -\sqrt{2} Y_2 & -\sqrt{2} Y_4
      \end{array}\right)$
    \\
    \bottomrule
  \end{tabular}
  \caption{{Relevant matrices in the neutrino sector of the superpotential in $\Gamma_5$
    models.}}
  \label{tab:A5-W-matrix}
\end{table}

The charged-lepton sector $w_e$ of the superpotential is
\begin{equation}
  w_e
  =
  \alpha {(E^c L)}_\1 \chi^2
  + \beta {(E^c L)}_\3 \chi \varphi
  + \gamma {(E^c L)}_\5 {(\varphi^2)}_\5
  + \delta {(E^c L)}_\1 {(\varphi^2)}_\1
  \equiv
  -E^{cT} \cY_e L~~~.
\end{equation}
{In what follows we set the flavons to their vevs and denote them by $\chi$, $\varphi_i$.
We absorb $\chi\ne 0$, $\varphi_1\ne 0$ and the Lagrangian parameter $\delta$
into $\alpha$, $\beta$, $\gamma$, $\varphi_2$ and $\varphi_3$. Once this is
done, the matrix $\cY_e$ takes the form}
\begin{equation}
  \cY_e
  =
  \left(
    \begin{array}{ccc}
      \alpha + 4 \gamma (1 - \varphi_2 \varphi_3) &
      (\beta + 6 \gamma) \varphi_3 &
      (-\beta + 6 \gamma) \varphi_2 \\
      (-\beta + 6 \gamma) \varphi_3 &
      6 \gamma \varphi_3^2 &
      \alpha + \beta - 2 \gamma (1 - \varphi_2 \varphi_3) \\
      (\beta + 6 \gamma) \varphi_2 &
      \alpha - \beta - 2 \gamma (1 - \varphi_2 \varphi_3) &
      6 \gamma \varphi_2^2
    \end{array}
  \right)~~~.
\end{equation}
{The charged lepton mass matrix $m_e$ has the same form as in Eq.~\eqref{me}, $m_e= {\cal Y}_e v  \cos\hat\beta/\sqrt{2}$.
Setting $\varphi_2 = \varphi_3 = 0$ and switching the last
two rows gives a diagonal $m_e$, with eigenvalues
\begin{equation}
  m_a = (\alpha + 4 \gamma)\frac{v}{\sqrt{2}} \cos\hat\beta~~~,
  \qquad
  m_b = (\alpha - \beta - 2 \gamma)\frac{v}{\sqrt{2}} \cos\hat\beta~~~,
  \qquad
  m_c = (\alpha + \beta - 2 \gamma)\frac{v}{\sqrt{2}} \cos\hat\beta~~~.
\end{equation}
As for the $\Gamma_4$ case, we treat $\tau$ and the VEVs $\varphi_{2,3}$ as parameters to be freely varied in our fit. The remaining parameters are the overall scale $\Lambda$ and the dimensionless constants $\alpha$, $\beta$ and $\gamma$. By enforcing $CP$ conservation, the latter three are required to be real. The dependence on $\tan\hat\beta$ can be absorbed into these parameters.}

\section{Results}
{In this section we identify which scenarios we analyse, state the experimental data used and report the results
of a chi-square analysis with the predictions of the models. 
In table \ref{tab:scenarios}, we list the seven scenarios which reproduce the data well, with a reasonable $\chi ^{2}_\textrm{min}$ and minimum number of parameters.}
{We will present results only for these scenarios, omitting those presenting a high $\chi ^{2}_\textrm{min}$ or a large number of parameters.}
We identify the different cases with a code referring to the modular level $\Gamma_4 \equiv S_4$ or $\Gamma_5 \equiv A_5$ ``4 (5)''; Weinberg or Seesaw ``W (S)''; $CP$ conserving or violating ``C (V)''. For the $A_5$ Weinberg scenario, we add the transformation property of the lepton triplet, whether this transforms as a \textbf{3}, or \textbf{3'} ``3 (3p)''.

We present the results in this section for which $\tau$ is not restricted to be in the fundamental domain, $|\textrm{Re}(\tau)|\le 1/2$, $|\tau|\ge 1$. However, in appendix \ref{appFundamental} we also include a full list of modular transformations to the set of input parameters which transforms $\tau$ into the fundamental region, as well as the explicit numerical values for these transformed parameters, which will yield the same set of physical observables. In this main text we list the non-fundamental region input parameters to avoid confusion stemming from spurious additional imaginary parameters which are just an artefact of a basis transformation.

\begin{table}[h]
\footnotesize
\begin{center}
\noindent\makebox[\textwidth]{
\begin{tabular}{c|c|c|c|c}
Model & Operator & $CP$ conservation & Charged Lepton sector & Case Identifier \\ \hline

$S_4$ & Weinberg & $\cancel{CP}$ & Diagonal & 4WV \\
$S_4$ & Seesaw & $\cancel{CP}$ & Diagonal & 4SV \\ 
$S_4$ & Weinberg & $CP$ & Modified & 4WC \\
$S_4$ & Seesaw & $CP$ & Modified & 4SC \\
\hline
$A_5$ & Weinberg, $\rho_L =\3$ & ${CP}$ & Modified & 5WC3 \\
$A_5$ & Weinberg, $\rho_L =\tp$ & ${CP}$ & Modified & 5WC3p \\
\multirow{ 2}{*}{$A_5$} & Seesaw,  & \multirow{ 2}{*}{${CP}$} & \multirow{ 2}{*}{Modified} & \multirow{ 2}{*}{5SC} \\
&$\rho_L =\3$, $\rho_N =\tp$, Im$(\varphi_{2,3})=0$&&&\\
\end{tabular}}
\caption{A list of the seven scenarios presented with good fits to data.}
\label{tab:scenarios}
\end{center}
\end{table}

\subsection{Fit to Leptonic Data}
\label{CSE}
In Tab. \ref{tabdata}, we list the experimental data and errors we use to calculate our pulls and $\chi^{2}_\textrm{min}$ values. For the Yukawa couplings, we use the renomalised values at $m_Z$ scale, as detailed in Ref. \cite{Antusch:2013jca}. For the neutrino oscillation data, we use the most recent results from the NuFit collboration, Ref. \cite{Esteban:2018azc}. For the calculation of our $\chi^{2}_\textrm{min}$, we assume the conservative estimate of gaussian errors, unless explicitly stated otherwise. {Even though current data seem to prefer normal to inverted neutrino mass ordering, we do not weight this option in our $\chi^2$ function.}

We show our results for all the considered $\Gamma _4$ and $\Gamma _5$ cases in the three tables contained in Tab. \ref{tab:S4scenarios} and \ref{tab:A5scenarios} respectively. For each case we present the point in parameter space which minimises the $\chi^{2}$, as a result of a numerical minimisation procedure. In the first table, one finds the predictions and, in parentheses, pulls to the six observed neutrino parameters: the two mass squared differences, $\Delta m_{21} ^{2} ,~ \Delta m_{3l}^{2}$ (where the latter refers to $\Delta m _{32} ^{2} >0 $ for NO and $\Delta m_{31} ^{2} <0$ for IO), three PMNS angles, $\theta_{12},~\theta_{13},~\theta_{23}$, and $CP$ violating phase, $\delta$; as well as the final $\chi^{2} _\textrm{min}$. In the second table, we list the predictions for each scenario for the: three individual neutrino masses, $m_1,~m_2,~m_3$; Majorana phases $\alpha_{21},~\alpha_{31}$; neutrinoless double beta decay parameter, $m_{ee}$; and Mass Ordering (MO). In the third table we specify the input parameters used to generate the best fit point discussed. In neither $\Gamma _4$, nor $\Gamma _5$ do we present the pulls from the Yukawa of the charged lepton sector, as we find sufficient freedom for every considered case to reproduce the observed values with negligible pulls $(\Delta \chi ^{2} <0.01)$.

\begin{table}[h!] 
\footnotesize
\begin{center}
\begin{tabular}{|c|c|}
\hline
$y_e(m_Z)$&$2.794745(16)\times 10^{-6}$\rule[-2ex]{0pt}{5ex}\\
\hline
$y_\mu(m_Z)$&$5.899863(19)\times 10^{-4}$\rule[-2ex]{0pt}{5ex}\\
\hline
$y_\tau(m_Z)$&$1.002950(91)\times 10^{-2}$\rule[-2ex]{0pt}{5ex}\\
\hline
\end{tabular}~~~~~~~~~~
\begin{tabular}{|c|c|c|}
\hline
&IO&NO\rule[-2ex]{0pt}{5ex}\\
 \hline
$\frac{\Delta m^2_{21}}{10^{-5}~{\rm eV}^2}$&$7.39(21)$&$7.39(21)$\rule[-2ex]{0pt}{5ex}\\
\hline
$\frac{\Delta m^2_{3\ell}}{10^{-3}~{\rm eV}^2}$&$-2.512(33)$&$+2.525(32)$\rule[-2ex]{0pt}{5ex}\\
\hline
$\sin^2\theta_{12}$&$0.310(13)$ &$0.310(13)$\rule[-2ex]{0pt}{5ex}\\
\hline
$\sin^2\theta_{13}$&$0.02263(66)$ &$0.02240(66)$\rule[-2ex]{0pt}{5ex}\\
\hline
$\sin^2\theta_{23}$&$0.582(17)$ &$0.582(17)$\rule[-2ex]{0pt}{5ex}\\
\hline
$\delta/\pi$&$1.56(15)$ &$1.21(19)$\rule[-2ex]{0pt}{5ex}\\
\hline
\end{tabular}
\caption{Left panel: charged lepton Yukawa couplings renormalized at the $m_Z$ scale, from Ref. \cite{Antusch:2013jca}. Right panel: neutrino oscillation data, from Ref. \cite{Esteban:2018azc}. The squared mass difference
$\Delta m^2_{3\ell}$ is equal to $\Delta m^2_{31}$ for normal ordering and $\Delta m^2_{32}$ for inverted ordering.
Errors, shown in brackets, {are the average of positive and negative 1$\sigma$ deviations. The $\chi^2$ function is not gaussian along the $\sin^2\theta_{23}$ direction and our definition overestimates the error.}}
\label{tabdata}
\end{center}
\end{table}
\newpage

\subsection{Numerical results at level 4}
\label{RS4}

To minimize the number of effective parameters, we first analyze the case of diagonal charged lepton sector. This can be realized by
fixing the VEV of the flavon $\varphi$ along the direction $(0,\varphi_2,0)$. All terms depending on $a'$ drop. The remaining input parameters $a$, $b$ and $c$, can be fixed to exactly reproduce the charged lepton masses:
\be
(a,b,c)=\frac{\sqrt{2}}{v\cos\beta}\left(\frac{m_e}{\varphi_2^3},\frac{m_\mu}{\varphi_2^2},\frac{m_\tau}{\varphi_2}\right)~~~.
\ee
Due to the hierarchical pattern in powers of the VEV, these input parameters may be all of similar order by fixing, for example, $|\varphi_2|=1/100$, which leads to
{
\be
a\cos\beta\simeq 2.8, \hspace{1cm}
b\cos\beta\simeq 5.9, \hspace{1cm}
c\cos\beta \simeq 1.
\ee
}
We are left with 3 Lagrangian parameters, ($\Lambda,{\tt Re}(\xi),{\tt Im}(\xi))$ and the (complex) modulus VEV $\tau$. 
Choosing the neutrino mass generated by the Weinberg operator (denoted case ``4WV''), we get a good agreement between the model and the data by the parameter choice shown in Tab. \ref{tab:S4scenarios}, with a $\chi^2 _\textrm{min} \sim 0.6$.
We also present results for the same scenario, but now with neutrino mass generated by a type-I seesaw (denoted case ``4SV''), with a $\chi^{2}_\textrm{min} \sim 1.1$.

We may further reduce the number of free parameters by imposing that the Lagrangian be $CP$ conserving. This amounts, in our basis, to requiring real Lagrangian parameters, i.e. Im$(\xi)=0$. We found no feasible solutions with this further restriction keeping the charged lepton sector diagonal as before. Relaxing this requirement, and setting $a'=0$, we find a good fit to data allowing small perturbations (in units of $\varphi _2$) of Im$(\varphi_1)=-$Im$(\varphi_3)\neq 0$. We present our results for this scenario for both the Weinberg case (denoted ``4WC''), with $\chi ^{2} _\textrm{min} \sim 3.2$ and the seesaw case (denoted ``4SC''), with $\chi^{2} _\textrm{min} \sim 0.3$. In both $CP$ conserving and violating scenarios, neutrino masses from the Weinberg operator have inverted ordering, while those coming from the seesaw mechanism are normal ordered.

\begin{table}[h!]
\footnotesize
\begin{center}
\noindent\makebox[\textwidth]{
\begin{tabular}{|c|c|c|c|c|c|c|c|}
\hline
&\multicolumn{6}{c|}{\tt value (pull)}&
\rule[-2ex]{0pt}{5ex}\\
\hline
Case &  $\Delta m^2_{21} \cdot 10^5~{\rm eV}^{-2}$& $\Delta m^2_{3l} \cdot 10^3~{\rm eV}^{-2}$ & $\sin ^{2} \theta_{12}$ &$\sin ^{2} \theta_{13}$ &$\sin ^{2} \theta_{23}$ &$\delta / \pi$ & $\chi ^{2} _{\textrm{min}}$
\rule[-2ex]{0pt}{5ex}\\
\hline
$4WV$ & 7.39 (0) & -2.517 (-0.2) & 0.310 (+0.0) & 0.02262 (-0.0)& 0.583 (+0.1)& 1.68 (+0.8) & 0.6
\rule[-2ex]{0pt}{5ex}\\
\hline
$4SV$ & 7.39 (0) & 2.527 (+0.1) & 0.310 (+0.0) & 0.02241 (+0.0)& 0.580 (-0.1)& 1.40 (+1.0) & 1.1
\rule[-2ex]{0pt}{5ex}\\
\hline
$4WC$ & 7.39 (0) & -2.512 (-0.0) & 0.310 (+0.0) & 0.02264 (+0.0)& 0.580 (-0.1)& 1.83 (+1.8) & 3.2
\rule[-2ex]{0pt}{5ex}\\
\hline
$4SC$ & 7.39 (0) & 2.526 (+0.0) & 0.317 (+0.5) & 0.02237 (-0.1)& 0.580 (-0.1)& 1.25 (+0.2) & 0.3
\rule[-2ex]{0pt}{5ex}\\
\hline
\end{tabular}}
\vspace{1cm}

\noindent\makebox[\textwidth]{
\begin{tabular}{|c|c|c|c|c|c|c|c|}
\hline
&\multicolumn{7}{c|}{\tt value}
\rule[-2ex]{0pt}{5ex}\\
\hline
Case &  $m_1 \cdot 10^2~{\rm eV}^{-2}$& $m_2 \cdot 10^2~{\rm eV}^{-2}$ & $m_3 \cdot 10^2~{\rm eV}^{-2}$ & $\alpha_{21} /\pi$ & $\alpha_{31} /\pi$ & $m_{ee} \cdot 10^2~{\rm eV}^{-1}$ & MO
\rule[-2ex]{0pt}{5ex}\\
\hline
$4WV$ & 6.56 & 6.61 & 4.31 & 0.21 & 1.76 & 6.18 & IO
\rule[-2ex]{0pt}{5ex}\\
\hline
$4SV$ & 4.23 & 4.32 & 6.57 & 0.22 & 0.54 & 4.01 & NO
\rule[-2ex]{0pt}{5ex}\\
\hline
$4WC$ & 6.33 & 6.39 & 3.96 & 1.88 & 1.69 & 6.20 & IO
\rule[-2ex]{0pt}{5ex}\\
\hline
$4SC$ & 4.26 & 4.35 & 6.59 & 0.11 & 0.30 & 4.25 & NO
\rule[-2ex]{0pt}{5ex}\\
\hline
\end{tabular}}

\vspace{1cm}

\noindent\makebox[\textwidth]{
\begin{tabular}{|c|c|c|c|c|c|c|c|c|c|}
\hline
&\multicolumn{9}{c|}{\tt Input parameters}
\rule[-2ex]{0pt}{5ex}\\
\hline
Case &  Re($\tau$) & Im$(\tau)$ & Re($\xi$) & Im($\xi$) & Im($\varphi_3$)=-Im($\varphi_2$)& $a $ & $b $ & $c $ & 1/$\Lambda$ (eV$^{-1}$)
\rule[-2ex]{0pt}{5ex}\\
\hline
$4WV$ & 1.155 & 0.9797 & -2.536 & -0.07654 & - & 2.795 & 5.900 & 1.003 & 0.007395
\rule[-2ex]{0pt}{5ex}\\
\hline
$4SV$  & 0.8436 & 0.9968 & -2.600 & 0.1151 & - & 2.795 & 5.900 & 1.003 & 0.7672
\rule[-2ex]{0pt}{5ex}\\
\hline
$4WC$  & 2.530 & 0.5380 & -0.1063 & - & -0.001063 & 2.647 & 5.899 & 0.9918 & 0.003799
\rule[-2ex]{0pt}{5ex}\\
\hline
$4SC$  & 2.506 & 0.5905 & -2.595 & - & 0.001081 & 2.642  & 5.899 & 0.9914 & 1.301
\rule[-2ex]{0pt}{5ex}\\
\hline
\end{tabular}}
\caption{Results of the fit to lepton data for the $\Gamma_4$ models. In the top panel, best values and pulls for the observables used in the fit. Also the minimum $\chi^2$ is shown.
In the middle table, predictions of the models: neutrino masses, phases and parameter $m_{ee}$ relevant for neutrinoless double beta decay. In the bottom panel input parameters at the minimum of the $\chi^2$ function. We have fixed $\varphi_2 =0.01$ for all four cases. {To simplify the notation, the factors $\cos\hat\beta$ and $1/\sin^2\hat\beta$ have been omitted from $a$, $b$, $c$ and $\Lambda$, respectively.}} \label{tab:S4scenarios}
\end{center}
\end{table}

In our setup we were unable to describe both the neutrino masses and the mixing matrix with fewer than five parameters. On the other hand the overall results and predictions are quite stable with respect to the details of the model. The quality of the fit is quite similar in all cases analysed and the results mainly depend on the choice between the Weinberg operator and the seesaw mechanism. In both cases the neutrino mass spectrum is nearly degenerate and the lightest neutrino mass is around 40 meV. When we adopt the Weinberg operator (seesaw mechanism) $m_{ee}$ is close to 60  (40) meV. A normally ordered spectrum (corresponding to the seesaw mechanism) predicting a relatively high $m_{ee}$ parameter seems a common feature
to most of the models enjoying modular invariance and providing a good fit to the data. The neutrino masses in our model are slightly heavier than those of the level 4 models studied in Ref. \cite{Penedo:2018nmg,Novichkov:2018ovf}.

\subsection{Numerical results at level 5}
\label{RA5}
{We now turn to the models at level 5. Unlike in level 4, all the examples listed here produce a $CP$ conserving Lagrangian. In our basis, this requirement is that all Lagrangian parameters be real. The charged lepton masses are essentially controlled by $\alpha, \beta, \gamma$, while neutrino masses and mixing angles are mainly
governed by $\Lambda$, $\tau$ and $\varphi$. We fix $\varphi _1 = 1$ and, to reduce the number of parameters,
we restrict the two VEVs of $(\varphi_2 , \varphi_3 )$ to real values. Neutrino
properties are thus described by a total of five parameters. 

As we can see from Tab.  \ref{tab:A5scenarios}, we get the best agreement with data when neutrino masses come from the Weinberg operator, with $\rho_L \sim \3$ (denoted case ``5WC3''), for which we get a $\chi ^{2} _\textrm{min} \sim 1.1$. 
The $\tau$ value is very close to the border of the fundamental region (see also Tab. \ref{tab:A5Fundamental} in Appendix C), where $CP$ is conserved. This result strongly supports the indication that, in a $CP$ invariant model, even a tiny departure from the region 
of moduli space where $CP$ is preserved can cause large observable $CP$-violating effects \cite{Novichkov:2019sqv}.
We also notice that all the components of the multiplet $\varphi$ are of the same order, indicating that the
charged lepton mass matrix is far from the diagonal form, related to $\varphi\propto (1,0,0)$. This is a new feature, since in the level 4 models discussed here and in the level 3 model of Ref. \cite{Criado:2018thu}, the contribution to the lepton mixing of the charged lepton mass matrix (depending on ordinary flavons) is small. The model predicts $m_{ee}\approx 27$ meV.
The mass ordering is inverted, as in all previous cases
dealing with the Weinberg operator. An exception is provided by the other Weinberg case at level 5 in which $\rho_L \sim \tp$ (denoted ``5WC3p''), which predicts normal ordering at the price of a considerably worse $\chi^{2} _\textrm{min} \sim 12.6$. The largest pulls are the one in $\delta$, which deviates by more than 3$\sigma$ and by $\sin^2\theta_{13}$,
about $1\sigma$ below the current best value.

We have also explored this model in a seesaw scenario, in which $\rho_L \sim \3 ,~ \rho_{N} \sim \tp$ (denoted ``5SC'').
The agreement with data is not excellent and our estimate of the $\chi^{2} _\textrm{min}$ is $11.1$. The main contributions to the $\chi^{2} _\textrm{min}$ come from $\delta$, which deviates by more than 2$\sigma$ and by $\sin^2\theta_{23}$,
about $2\sigma$ below the current best value. For $\sin ^{2} \theta_{23} \simeq 0.45$ we do not use the nominal 
pull, since the error is non-gaussian. We assess the contribution to the $\chi^{2} _\textrm{min}$ directly using the results from NuFit. The neutrino mass spectrum has normal ordering. Specific to the seesaw realization are the prediction of $\theta_{23}$ in the first octant and of a vanishing $m_1$. The latter result has no counterpart in any model 
based on modular invariance so far investigated. As a consequence $m_{ee}\approx 1.3$ meV is rather small.


\begin{table}[h!]
\footnotesize
\begin{center}
\noindent\makebox[\textwidth]{
\begin{tabular}{|c|c|c|c|c|c|c|c|}
\hline
&\multicolumn{6}{c|}{\tt value (pull)}&
\rule[-2ex]{0pt}{5ex}\\
\hline
Case &  $\Delta m^2_{21} \cdot 10^5~{\rm eV}^{-2}$& $\Delta m^2_{3l} \cdot 10^3~{\rm eV}^{-2}$ & $\sin ^{2} \theta_{12}$ &$\sin ^{2} \theta_{13}$ &$\sin ^{2} \theta_{23}$ &$\delta / \pi$ & $\chi ^{2} _{\textrm{min}}$
\rule[-2ex]{0pt}{5ex}\\
\hline
$5WC3$ & 7.39 (0) & -2.512 (+0.0) & 0.312 (+0.1) & 0.02260 (-0.0)& 0.592 (+0.6)& 1.69 (+0.9) & 1.1
\rule[-2ex]{0pt}{5ex}\\
\hline
$5WC3p$ & 7.39 (0) & 2.525 (+0.0) & 0.309 (-0.1) & 0.0217 (-1.2)&  0.586 (+0.3)&  0.57 (-3.3) & 12.6
\rule[-2ex]{0pt}{5ex}\\
\hline
$5SC$ & 7.39 (0) & 2.522 (-0.1) & 0.292 (-1.4) & 0.0228 (+0.5) & 0.449 (-2.0*)&  1.63 (+2.2) & 11.1*
\rule[-2ex]{0pt}{5ex}\\
\hline
\end{tabular}}
\vspace{1cm}

\noindent\makebox[\textwidth]{
\begin{tabular}{|c|c|c|c|c|c|c|c|}
\hline
&\multicolumn{7}{c|}{\tt value}
\rule[-2ex]{0pt}{5ex}\\
\hline
Case &  $m_1 \cdot 10^2~{\rm eV}^{-2}$& $m_2 \cdot 10^2~{\rm eV}^{-2}$ & $m_3 \cdot 10^2~{\rm eV}^{-2}$ & $\alpha_{21} /\pi$ & $\alpha_{31} /\pi$ & $m_{ee} \cdot 10^2~{\rm eV}^{-1}$ & MO
\rule[-2ex]{0pt}{5ex}\\
\hline
$5WC3$ & 4.94 & 5.01 & 0.0942 & 0.70 & 0.94 & 2.7 & IO
\rule[-2ex]{0pt}{5ex}\\
\hline
$5WC3p$ & 2.82 & 2.95 & 5.76 & 0.38 & 0.26 & 2.3 & NO
\rule[-2ex]{0pt}{5ex}\\
\hline
Case &  $m_1 \cdot 10^2~{\rm eV}^{-2}$& $m_2 \cdot 10^2~{\rm eV}^{-2}$ & $m_3 \cdot 10^2~{\rm eV}^{-2}$ & \multicolumn{2}{|c|}{$(\alpha_{21} -\alpha_{31}) /\pi$} & $m_{ee} \cdot 10^2~{\rm eV}^{-1}$ & MO
\rule[-2ex]{0pt}{5ex}\\
\hline
$5SC$ & 0 & 0.860 & 5.02 & \multicolumn{2}{|c|}{1.68} & 0.13 & NO
\rule[-2ex]{0pt}{5ex}\\
\hline
\end{tabular}}
\vspace{1cm}

\noindent\makebox[\textwidth]{
\begin{tabular}{|c|c|c|c|c|c|c|c|c|c|c|}
\hline
&\multicolumn{10}{c|}{\tt Input parameters}
\rule[-2ex]{0pt}{5ex}\\
\hline
Case &  Re($\tau$) & Im$(\tau)$ & Re($\varphi_2$) & Im($\varphi_2$) & Re($\varphi_3$) & Im($\varphi_3$) & $\alpha \cdot 10^{3}$ & $\beta \cdot 10^{3}$ & $\gamma \cdot 10^{3}$ & 1/$\Lambda$ (eV$^{-1}$)
\rule[-2ex]{0pt}{5ex}\\
\hline
$5WC3$ & -0.01882 & 0.9929 & 0.4260 & - & 0.8030 & - & 3.018 & 3.927 & -0.4484 & 0.008180
\rule[-2ex]{0pt}{5ex}\\
\hline
$5WC3p$ & -0.09033 & 0.2190 & 0.4244 & - & 0.01694 & - & 3.259 & 4.311 & -0.8036 & 0.0006303
\rule[-2ex]{0pt}{5ex}\\
\hline
$5SC$  & -0.3615 & 0.2412 & 0.04759 & - & 0.3731 & - & 3.368 & 4.411 & -0.8126 & 0.0001639
\rule[-2ex]{0pt}{5ex}\\
\hline
\end{tabular}}
\caption{Results of the fit to lepton data for the $A_5$ models. For the $5SC$ case, the predicted lightest neutrino mass is $m_1=0$ and so only one physical Majorana phase exists, which appears in the combination $(\alpha_{21}-\alpha_{31})$ in neutrinoless double beta decay and hence we report only this combination. We have fixed $\varphi_1 =1$ for all three cases. *Actual NuFit 4.0 error on $\sin ^{2} \theta _{23}$ measurement (for NO) used, rather than assumed Gaussian error.  {To simplify the notation, the factors $\cos\hat\beta$ and $1/\sin^2\hat\beta$ have been omitted from $\alpha$, $\beta$, $\gamma$ and $\Lambda$, respectively.}}
\label{tab:A5scenarios}
\end{center}
\end{table}

In all these cases we find that the spread of the parameters $\alpha$, $\beta$, $\gamma$ is less than one order of magnitude, much less than the one among the charged lepton masses. Our approach and the related results significantly differ from those of refs. \cite{Novichkov:2018nkm,Ding:2019xna} where
several modular invariant models at level 5 have been analysed, under the assumption that
the charged lepton sector be always diagonal \cite{Novichkov:2018nkm} or diagonal when
depending on ordinary flavons \cite{Ding:2019xna}. We have also looked for a better agreement with data in the seesaw case by relaxing the requirement of a real $(\varphi_2 , \varphi _3)$. At the price of more parameters, we obtain an better fit to data, though we do not present this example explicitly.}

\section{Conclusion}
Modular invariance have been proven to offer a promising framework to describe lepton masses and mixing angles. In minimal models masses, mixing angles and phases are all predicted in terms of the modulus in addition to a few free parameters. Despite these nice features, neutrinos and charged leptons
typically require different realizations to reproduce the sizeable hierarchy among electron, muon and tau masses. In most of the existing models
right-handed leptons are assign to singlets of the modular group to allow a sufficient number of free parameters, tuned to match the charged lepton masses.
We think that this aspect might indicate the need for a different description, perhaps in terms of other moduli than the one controlling the neutrino sector.
In a simple-minded approach, not aiming at a fundamental description but rather to test the ground for a more extensive analysis, we have explored
alternative realisations of the charged lepton sector in modular invariant models at levels 4 and 5.

At level 4 we have shown that it is possible to ascribe the charged lepton mass hierarchy to the weight difference in the right handed sector,
similar to what occurs in Froggatt-Nielsen models, wherein the role of the weights is played by the charges. At level 5 we have assigned both right-handed and left-handed leptons to
irreducible triplets of the finite modular group $\Gamma_5$. {Moreover we have shown that also at level 5 the three parameters required to describe charged lepton masses can be almost within the same order of magnitude. In all models considered here
we do not need a strong hierarchy at the level of Lagrangian parameters to reproduce charged lepton masses.

We built several models along these lines, analysing neutrino masses coming either from the Weinberg operator or from a type I seesaw, and we have selected seven scenarios which produce a reasonable fit to data, four of them at level 4 and three at level 5. We looked for minimal realisations, in terms of the lowest possible number of free parameters. 
Among them we also count the vacuum expectation values of both modulus and flavons, which we varied in order to maximise the agreement with the data. Three parameters are in a one-to-one relation with the charged lepton masses.
Besides them, all of our scenarios make use of five parameters, always including an overall scale $\Lambda$, 
and real and imaginary parts of $\tau$. In these cases we get four predictions: the absolute neutrino mass scale
and all $CP$ violating phases, which allow one to pin down the value of $m_{ee}$, relevant to neutrinoless double beta decay.
So far few models based on modular invariance perform better, managing to fit the neutrino data with four free
parameters. 
In all cases analysed at level 5 and in two cases at level 4 we demanded that the Lagrangian be $CP$ conserving.
A common feature of level 4 and 5 scenarios is that inverted ordering for neutrino masses is predicted when adopting
the Weinberg operator and normal ordering when making use of type I seesaw, with a single exception whose $\chi^{2} _\textrm{min}$ is not particularly good. 
At level 4 the overall results and predictions are quite stable with respect to the details of the model, only depending
on the choice between the Weinberg operator and the seesaw mechanism. In both cases the neutrino mass spectrum is nearly degenerate and the lightest neutrino mass is around 40 meV. At level 5 we get an excellent $\chi^{2} _\textrm{min}$
only when considering neutrino masses generated by the Weinberg operator, predicting inverted mass ordering. In the seesaw scenario a good fit requires the introduction of additional parameters. Remarkably we find that our seesaw models at level 5 predict a massless neutrino.}

We do not consider our results conclusive and we think that there is still a considerable room to improve the characterization of the charged lepton sector.
Nevertheless, by exploring some nonstandard possibilities, we hope to have provided some 
new element for the identification of a basic framework.

\section*{Acknowledgements}
This project has received support in part by the MIUR-PRIN project 2015P5SBHT 003 ``Search for the Fundamental Laws and Constituents''
and by the European Union's Horizon 2020 research and innovation programme under the Marie Sklodowska-Curie grant 
agreement N$^\circ$~674896 and 690575. The research of F.~F.~was supported in part by the INFN.
The research of J. C. C. was supported by the Spanish MINECO project
FPA2016-78220-C3-1-P, the Junta de Andaluc\'ia grant FQM101 and the
Spanish MECD grant FPU14. SJDK would like to thank Gui-Jun Ding for useful discussions.

\newpage
\begin{appendices}
\setlength{\extrarowheight}{0.0 cm}

\section{Finite modular group $\Gamma_4$ and level 4 modular forms}
\label{appendix:level_4}

The finite modular group $\Gamma_4$ is isomorphic to $S_4$, the symmetric group of permutations of four objects. 
It has $24$ elements and five irreducible representations:
$\1$, $\op$, $\2$, $\3$ and $\tp$.
It admits a presentation in terms of two
generators $S$ and $T$:
\begin{align}
S^2 \,=\, (ST)^3 \,=\, T^4 \,=\, \mathds{1}\,.
\label{eq:present42}
\end{align}
In this paper we use an explicit realization of the elements $S$ and $T$ for the
different representations, obtained from the one in Ref.~\cite{Bazzocchi:2009pv}, with the identification \cite{Penedo:2018nmg}: $S=S'T'^2$ and $T=S'$, where the primed generators are those given in Ref.~\cite{Bazzocchi:2009pv}. We also use the Clebsch-Gordan coefficients listed in Ref.~\cite{Bazzocchi:2009pv}.

The linear space of weight 2 and level 4 modular forms has dimension 5
(see, e.g., \cite{Feruglio:2017spp}).
These forms can be constructed in terms of the Dedekind eta function ~\cite{Penedo:2018nmg}:
\be
\eta(\tau) \equiv q^{1/24} \prod_{n=1}^{\infty}\left(1 - q^n\right)\,,
\qquad
q = e^{2\pi i\tau}\,.
\ee
%
Defining
\begin{align}
Y(c_1,\dots,c_6|\tau) &\equiv \frac{\di}{\di\tau} \bigg[
c_1 \log \eta\left(\tau+\frac{1}{2}\right) +
c_2 \log \eta\left(4\tau\right) +
c_3 \log \eta\left(\frac{\tau}{4}\right) \nonumber\\
&+ c_4 \log \eta\left(\frac{\tau+1}{4}\right) + 
c_5 \log \eta\left(\frac{\tau+2}{4}\right) +
c_6 \log \eta\left(\frac{\tau+3}{4}\right)\bigg]\,,
\end{align}
%
with $c_1+\dots+c_6 = 0$, the basis of the modular forms of weight 2 reads \cite{Penedo:2018nmg},
{\begin{subequations}
\label{eq:S4_modular_forms_analytic}
\begin{align}
Y_1(\tau) &\equiv  i~ Y(1,1,\omega,\omega^2,\omega,\omega^2|\tau)\,,\\
Y_2(\tau) &\equiv i~ Y(1,1,\omega^2,\omega,\omega^2,\omega|\tau)\,,\\
Y_3(\tau) &\equiv i~ Y(1,{-1},{-1},{-1},1,1|\tau)\,,\\
Y_4(\tau) &\equiv i~ Y(1,-1,-\omega^2,-\omega,\omega^2,\omega|\tau)\,,\\
Y_5(\tau) &\equiv i~ Y(1,-1,-\omega,-\omega^2,\omega,\omega^2|\tau)\,,
\end{align}
\end{subequations}
}
%
with $\omega \equiv e^{2\pi i/3}$. {Notice here, we have an extra factor of $i$ compared to the definition of Ref. \cite{Penedo:2018nmg}.} It has been shown that $Y_1(\tau)$ and $Y_2(\tau)$ form a doublet transforming in the $\2$ of $S_4$, 
while the three remaining modular forms
make up a triplet transforming in $\tp$ of $S_4$.
Doublet and the triplet will be denoted by
\be
Y_\2(\tau) \equiv
\begin{pmatrix}Y_1(\tau)\\ Y_2(\tau)\end{pmatrix}\,,
\qquad
Y_\tp(\tau) \equiv
\begin{pmatrix}Y_3(\tau)\\ Y_4(\tau)\\ Y_5(\tau)\end{pmatrix}\,.
\ee
%
The $q$-expansions ($q\equiv e^{i 2\pi\tau}$) for Eq. \eqref{eq:S4_modular_forms_analytic} can be found in \cite{Penedo:2018nmg}. In our analysis we use the full analytic form.
The modular forms of higher weights $k = 4,\,6,\dots$ are homogeneous polynomials
in the variables $Y_i(\tau)$, $i=1,\dots,5$. 

\noindent
Under $CP$, Eq. \eqref{tauCP}, modular forms of level 4 and weight 2 transform as \cite{Novichkov:2018ovf}:
\be
Y_\2(-\tau^*)=X_\2 
\left[Y_\2(\tau)\right]^*~~~,~~~~~~~Y_\tp(-\tau^*)=X_\tp 
\left[Y_\tp(\tau)\right]^*~~~,
\ee
where $X_\2$ and $X_\tp$ are the matrices:
\be
X_\2=\begin{pmatrix}
0 & 1 \\ 1 & 0
\end{pmatrix}~~~,~~~~~~~
X_\tp=-\frac{1}{3}
\begin{pmatrix}
-1 & 2 \omega & 2 \omega^2 \\
 2 \omega & 2 \omega^2 & -1  \\
2 \omega^2 & -1 & 2 \omega  
\end{pmatrix}\,.
\label{CPMF}
\ee
By decomposing products of representations in their irreducible components we find that a consistent action of $CP$ on chiral multiplets $\varphi_{\mathbf{r}}$ transforming 
in the representation $\mathbf{r}$ $(\mathbf{r}=\1,\op,\2,\3,\tp)$ of $\Gamma_4$ is given by:
\be
\varphi_{\mathbf r} \xrightarrow{\text{$CP$}} X_{\mathbf r}~ \varphi_{\mathbf r}^*~~~,
\ee
with $X_\2$ and $X_\tp$ given above and
\be
X_\1=-X_\op=1~~~,~~~~~X_\3=X_\tp~~~.
\ee
This set of matrices satisfy the consistency conditions
\be
X_{\mathbf r}~ \rho^*_{\mathbf r}(\gamma)~ X_{\mathbf r}^{-1}= \rho_{\mathbf r}(\gamma')~~~,~~~~~~~~~~~(\gamma,\gamma')\in \Gamma~~~,
\ee
as can be checked by working with the generators $\gamma=(S,T)$. We find $S'=S^{-1}$ and $T'=T^{-1}$.
In our basis, the requirement of $CP$ conservation on a modular invariant supersymmetric theory at level 4, adopting the above $CP$
transformations on the chiral multiplets,
amounts to having all Lagrangian parameters real.
\section{Finite modular group $\Gamma_5$ and level 5 modular forms}
\label{app5}
The finite modular group $\Gamma_5$ is isomorphic to $A_5$, the group of even permutations of five objects. It has 60 elements and five irreducible representations: $\mathbf{1}$, $\mathbf{3}$, $\mathbf{3'}$, $\mathbf{4}$ and $\mathbf{5}$. It admits a presentation in terms of two generators of $S$ and $T$:
\begin{align}
S^2 \,=\, (ST)^3 \,=\, T^5 \,=\, I\,.
\label{eq:A5presentation}
\end{align}

\noindent
In this paper we use the explicit realisation of the elements $S$ and $T$ for the different representations given in Ref. \cite{Ding:2011cm}, where we can also find the corresponding
Clebsch-Gordan coefficients.

Level 5 modular forms of weight 2 have been built in Ref. \cite{Novichkov:2018nkm},  making use of the Jacobi theta function:
\be
\theta_3 (u,\tau) \equiv \theta _{0,0} (u,\tau) = \sum_{n=-\infty} ^{\infty} p^{n^{2}} e^{2 \pi i n u} ~~~,
\ee
where $p\equiv e^{\pi i \tau}$. Defining the seed functions:
\begin{equation}
\begin{aligned}[c]
\alpha_{1,-1}(\tau) \,&\equiv\, \theta_3\left( \frac{\tau+1}{2}, 5\tau\right)
\,, \\
\alpha_{1,0}(\tau) \,&\equiv\, \theta_3\left( \frac{\tau+9}{10}, \frac{\tau}{5}\right)
\,, \\
\alpha_{1,1}(\tau) \,&\equiv\, \theta_3\left( \frac{\tau}{10}, \frac{\tau+1}{5}\right)
\,, \\
\alpha_{1,2}(\tau) \,&\equiv\, \theta_3\left( \frac{\tau+1}{10}, \frac{\tau+2}{5}\right)
\,, \\
\alpha_{1,3}(\tau) \,&\equiv\, \theta_3\left( \frac{\tau+2}{10}, \frac{\tau+3}{5}\right)
\,, \\
\alpha_{1,4}(\tau) \,&\equiv\, \theta_3\left( \frac{\tau+3}{10}, \frac{\tau+4}{5}\right)
\,,
\end{aligned}
\qquad
\begin{aligned}[c]
\alpha_{2,-1}(\tau) \,&\equiv\, e^{2 \pi i \tau / 5} \, 
\theta_3\left( \frac{3\tau+1}{2}, 5\tau\right)
\,, \\
\alpha_{2,0}(\tau) \,&\equiv\, \theta_3\left( \frac{\tau+7}{10}, \frac{\tau}{5}\right)
\,, \\
\alpha_{2,1}(\tau) \,&\equiv\, \theta_3\left( \frac{\tau+8}{10}, \frac{\tau+1}{5}\right)
\,, \\
\alpha_{2,2}(\tau) \,&\equiv\, \theta_3\left( \frac{\tau+9}{10}, \frac{\tau+2}{5}\right)
\,, \\
\alpha_{2,3}(\tau) \,&\equiv\, \theta_3\left( \frac{\tau}{10}, \frac{\tau+3}{5}\right)
\,, \\
\alpha_{2,4}(\tau) \,&\equiv\, \theta_3\left( \frac{\tau+1}{10}, \frac{\tau+4}{5}\right)
\,,
\end{aligned}
\label{eq:seed}
\end{equation}
and the functions,
\begin{align}
Y(c_{1,-1},\ldots,c_{1,4};c_{2,-1},\ldots,c_{2,4}|\tau) \equiv \sum_{i,j} c_{i,j}
\frac{\di}{\di\tau}\log\alpha_{i,j}(\tau)
\,,\quad \textrm{with } \sum_{i,j} c_{i,j} = 0\,,
\end{align}
then the modular forms of weight two are divided into the following multiplets of $A_5$,
\begin{align}
Y_\mathbf{5}(\tau) = \left(\begin{array}{c}
Y_{1}(\tau)\\
Y_{2}(\tau)\\
Y_{3}(\tau)\\
Y_{4}(\tau)\\
Y_{5}(\tau)
\end{array}\right)
&\equiv\left(\begin{array}{c}
-\frac{1}{\sqrt{6}}Y\left(-5,1,1,1,1,1;-5,1,1,1,1,1\middle|\tau\right)\\
Y(0,1,\zeta^4,\zeta^3,\zeta^2,\zeta\,;\,0,1,\zeta^4,\zeta^3,\zeta^2,\zeta\,|\,\tau)\\
Y(0,1,\zeta^3,\zeta,\zeta^4,\zeta^2\,;\,0,1,\zeta^3,\zeta,\zeta^4,\zeta^2\,|\,\tau)\\
Y(0,1,\zeta^2,\zeta^4,\zeta,\zeta^3\,;\,0,1,\zeta^2,\zeta^4,\zeta,\zeta^3\,|\,\tau)\\
Y(0,1,\zeta,\zeta^2,\zeta^3,\zeta^4\,;\,0,1,\zeta,\zeta^2,\zeta^3,\zeta^4\,|\,\tau)
\end{array}\right)\,,
\label{eq:5} \\[2mm]
Y_\mathbf{3}(\tau) = \left(\begin{array}{c}
Y_{6}(\tau)\\
Y_{7}(\tau)\\
Y_{8}(\tau)
\end{array}\right)
&\equiv\left(\begin{array}{c}
\frac{1}{\sqrt{2}}Y\left(-\sqrt{5},-1,-1,-1,-1,-1;\sqrt{5},1,1,1,1,1\middle|\tau\right)\\
Y(0,1,\zeta^4,\zeta^3,\zeta^2,\zeta\,;\,0,-1,-\zeta^4,-\zeta^3,-\zeta^2,-\zeta\,|\,\tau)\\
Y(0,1,\zeta,\zeta^2,\zeta^3,\zeta^4\,;\,0,-1,-\zeta,-\zeta^2,-\zeta^3,-\zeta^4\,|\,\tau)
\end{array}\right)\,,
\label{eq:3} \\[2mm]
Y_\mathbf{3'}(\tau) = \left(\begin{array}{c}
Y_{9}(\tau)\\
Y_{10}(\tau)\\
Y_{11}(\tau)
\end{array}\right)
&\equiv\left(\begin{array}{c}
\frac{1}{\sqrt{2}}Y\left(\sqrt{5},-1,-1,-1,-1,-1;-\sqrt{5},1,1,1,1,1\middle|\tau\right)\\
Y(0,1,\zeta^3,\zeta,\zeta^4,\zeta^2\,;\,0,-1,-\zeta^3,-\zeta,-\zeta^4,-\zeta^2\,|\,\tau)\\
Y(0,1,\zeta^2,\zeta^4,\zeta,\zeta^3\,;\,0,-1,-\zeta^2,-\zeta^4,-\zeta,-\zeta^3\,|\,\tau)
\end{array}\right)\,,
\label{eq:3p}
\end{align}
where $\zeta = e^{2\pi i/5}$.
The first few terms of the $q$-expansions of these modular forms can be found in Ref. \cite{Novichkov:2018nkm}. Our numerical results have made use of $q$-expansions up to $\mathcal{O}(q^{100})$, but the results are unchanged when using up to $\mathcal{O}(q^{5})$.

\newpage
\section{Numerical Results in Fundamental Region}
\label{appFundamental}
{The models studied in this paper are modular invariant and it is always possible to map the Lagrangian referred to a certain value $\tau$ of the modulus to an equivalent Lagrangian where the modulus $\tau'$ is inside the fundamental region $|\textrm{Re}(\tau')|\le 1/2$, $|\tau'|\ge 1$. By definition there exists a modular transformation $\gamma$ such that $\tau'=\gamma\tau$.
Together with the transformation $\tau\to\gamma\tau$, we consider the field redefinition mapping all chiral multiplets
except $L$ into the modular transformed ones, after setting to zero all their weights. We find that the low-energy superpotential
\be
w=-\frac{v^2}{2\Lambda}L^T~ {\cal W}(\tau)~ L-\frac{v}{\sqrt{2}}E^{cT}~ {\cal Y}(\varphi)~ L
\ee
becomes
\be
w=-\frac{v^2}{2\Lambda}L^T~ {\cal W}(\gamma\tau)~ L-\frac{v}{\sqrt{2}} E^{cT}~ {\cal Y}(\varphi)~ \rho_L^\dagger(\gamma)L~~~,
\ee
where
\be
{\cal W}(\gamma\tau) = (c\tau +d)^{2} \rho_L (\gamma)^{*}~ {\cal W}(\tau)~ \rho_L^\dagger (\gamma) ~~~.
\ee
Neutrino and charged lepton mass matrices are now:
\be
m_\nu=\frac{v^2}{\Lambda}~ {\cal W}(\gamma\tau)~~~,~~~~~~~~m_e^\dagger m_e=\frac{v^2}{2}\rho_L(\gamma)~ {\cal Y}(\varphi)^\dagger{\cal Y}(\varphi)~ \rho_L^\dagger(\gamma)~~~.
\ee
The lepton mixing matrix is unchanged. We list here the transformations needed to map the values of $\tau$ found by our minimisation procedure
to points inside the fundamental region.}

\begin{table}[h!]
\noindent\makebox[\textwidth]{
\begin{tabular}{|c|c|c|c|c|c|c|c|c|c|c|c|}
\hline
&\multicolumn{10}{c|}{\tt Input parameters - fundamental region}
\rule[-2ex]{0pt}{5ex}\\
\hline
Case & $\gamma \tau$ &Re($\tau$) & Im$(\tau)$ & Re($\varphi_1$) & Im($\varphi_1$) & Re($\varphi_2$) & Im($\varphi_2$) & Re($\varphi_3$) & Im($\varphi_3$) & 
 1/$\Lambda$ (eV$^{-1}$)
\rule[-2ex]{0pt}{5ex}\\
\hline
$4WV$ & $S T^{-1} \tau$ & -0.1579 & 0.9957 & 2/3 & 0 & 1/6 & $1/2 \sqrt{3}$ & -1/3 & $1/\sqrt{3}$ & 0.003223
\rule[-2ex]{0pt}{5ex}\\
\hline
$4SV$ & $T^{-1} \tau$ &  -0.1564 & 0.9968 & -1/3 & -$1/\sqrt{3}$ & -1/3 & 1/$\sqrt{3}$ & -1/3 & 0 & 0.7672
\rule[-2ex]{0pt}{5ex}\\
\hline
$4WC$  &$T^{-1}S T^{-3}\tau$  & -0.07915 &1.055 & -0.3947 & 0.5774 & 0.6974 & -0.05315 & 0.1053 & 0.1824  & 0.0007030
\rule[-2ex]{0pt}{5ex}\\
\hline
$4SC$  & $T^{-1}ST^{-3} \tau$ &-0.1667&0.9966&-0.2709&0.5774&0.6355&0.05406&0.2291&0.3968&0.06993
\rule[-2ex]{0pt}{5ex}\\
\hline
\end{tabular}}
\label{tab:S4Fundamental}
\caption{{Parameters $\tau$ and $\varphi$ in the fundamental region for level 4 models.}}
\end{table}

\begin{table}[h!]
\noindent\makebox[\textwidth]{
\begin{tabular}{|c|c|c|c|c|c|c|c|c|c|c|c|}
\hline
&\multicolumn{10}{c|}{\tt Input parameters - fundamental region}
\rule[-2ex]{0pt}{5ex}\\
\hline
Case & $\gamma \tau$ &Re($\tau$) & Im$(\tau)$ & Re($\varphi_1$) & Im($\varphi_1$) & Re($\varphi_2$) & Im($\varphi_2$) & Re($\varphi_3$) & Im($\varphi_3$) & 
 1/$\Lambda$ (eV$^{-1}$)
\rule[-2ex]{0pt}{5ex}\\
\hline
$5WC3$ & $S \tau$ & 0.01908 & 1.007 & -0.3301 & 0 & -0.7188 & 0 & -1.096 & 0 & 0.007958
\rule[-2ex]{0pt}{5ex}\\
\hline
$5WC3p$ & $T^{-2}S \tau$ &  -0.3908 & 3.902 & -0.1618 & 0 & 0.1621 & 0.4990 & 0.2911 & -0.8960 & 0.0007302
\rule[-2ex]{0pt}{5ex}\\
\hline
$5SC$  &$T^{-2}S \tau$  & -0.08591 &1.277 & 0.1812 & 0 & 0.4561 & 0.3314 & 0.7194 &-0.5227  & 0.002804
\rule[-2ex]{0pt}{5ex}\\
\hline
\end{tabular}}
\caption{{Parameters $\tau$ and $\varphi$ in the fundamental region for level 5 models.}}
\label{tab:A5Fundamental}
\end{table}

\clearpage

\end{appendices}

\end{document}